\documentclass[showpacs,aps,pra,superscriptaddress,10pt]{revtex4-1}

\usepackage{amsmath}
\usepackage{bm}
\usepackage{ucs}
\usepackage{geometry}
\usepackage{graphicx}
\usepackage[caption=false]{subfig}
\usepackage{epstopdf} 
\usepackage{refcount}
\usepackage{array}
\usepackage{hyperref}
\usepackage[dvipsnames]{xcolor}

\begin{document}
\title{Crossover between snake instability and Josephson instability of dark solitons in superfluid Fermi gases}
\author{W. Van Alphen}
\email{wout.vanalphen@uantwerpen.be}
\affiliation{TQC, Universiteit Antwerpen, Universiteitsplein 1, B-2610 Antwerpen, Belgium}
\author{H. Takeuchi}
\affiliation{Department of Physics and Nambu Yoichiro Institute of Theoretical and Experimental Physics (NITEP), Osaka City University, Osaka 558-8585, Japan}
%\author{S. N. Klimin}
%\affiliation{TQC, Universiteit Antwerpen, Universiteitsplein 1, B-2610 Antwerpen, Belgium}
%\affiliation{Department of Theoretical Physics, State University of Moldova, 2009 Chi\unichar{537}in\unichar{259}u, Moldova}
\author{J. Tempere}
\affiliation{TQC, Universiteit Antwerpen, Universiteitsplein 1, B-2610 Antwerpen, Belgium}
\affiliation{Lyman Laboratory of Physics, Harvard University, Cambridge, Massachusetts 02138, USA}
\begin{abstract} 
Dark solitons in superfluid Bose gases decay through the snake instability mechanism, unless they are strongly confined. Recent experiments in superfluid Fermi gases have also interpreted soliton decay via this mechanism. However, we show using both an effective field numerical simulation and a perturbative analysis that there is a qualitative difference between soliton decay in the BEC- and BCS-regimes. On the BEC-side of the interaction domain, the characteristic snaking deformations are induced by fluctuations of the amplitude of the order parameter, while on the BCS-side, fluctuations of the phase destroy the soliton core through the formation of local Josephson currents. The latter mechanism is qualitatively different from the snaking instability and this difference should be experimentally detectable.
\end{abstract} 
\maketitle
\section{Introduction}
\label{sec:intro}
Arising from an interplay between nonlinear and dispersive effects, solitons are solitary matter waves which retain their shape while propagating at a constant velocity. They emerge in a wide variety of physical systems including optical fibers, classical fluids, plasmas and ultracold atomic gases. Due to their high experimental tunability, ultracold atom clouds in particular form an ideal environment for studying the physics of nonlinear matter waves in a well-controlled way. In these atomic clouds, solitons often manifest themselves as \emph{dark} solitons, which are characterized by a localized density dip and a jump in the phase profile of the order parameter. Dark solitons have been theoretically and experimentally studied in Bose-Einstein condensates (BECs) \cite{THFrantzeskakis,EXPDenschlag,EXPBurgerBongs,EXPAndersonHaljian,EXPBeckerStellmer},  as well as superfluid Fermi gases \cite{THAntezzaDalfovo,THScottDalfovo,THLiaoBrand}. 
In both systems, they are subject to an instability mechanism called the snake instability \cite{EXPKu2,THMuryshev,THBrandReinhardt,THCetoliBrand}, which makes the soliton decay into vortices if the radial width of the atom cloud is too large. The name ``{snake instability}'' comes from the observation that in the decay process the depletion plane of the soliton starts to oscillate until it breaks up into one or more vortex structures. The snaking mechanism and the subsequent decay of the dark soliton plane have been observed experimentally in both BECs\cite{EXPDonadello} and superfluid Fermi gases \cite{EXPKu2}. 

Various theoretical methods have been used to analyze the instability of dark solitons in fermionic systems \cite{THCetoliBrand,THMunozBrand,THLvAKTSI}. In Ref.\ \cite{THLvAKTSI}, the critical length scale of the instability was calculated in the BEC-BCS crossover by means of a recently developed low-energy effective field theory (EFT) \cite{THKTLDEpjB}. This formalism, which is capable of describing Fermi superfluids in a wide range of temperatures and interaction strengths, is based on the assumption that the order parameter of the system changes slowly in both space and time. It has already been successfully employed for the description of the properties and dynamics of dark solitons as a function of temperature and population imbalance \cite{THKTDPrA,THLvAKTPrA,THvALKTColl}.

In the present paper, numerical simulations based on the EFT reveal that the dynamics of the dark soliton decay change significantly across the interaction domain. A perturbative analysis of the amplitude and phase fluctuations of the order parameter demonstrates that this crossover in the instability dynamics is related to a change in the nature of the unstable modes, which shift from amplitude-like to phase-like when one moves from the BEC- to the BCS-regime. All calculations are preformed for the case of a Fermi superfluid with a uniform background.  While traditionally ultracold gases are studied in set-ups with harmonic trapping potentials, the recent realization of box-like optical traps \cite{EXPGauntSchmidutz} provides an incentive to investigate uniform superfluids and the opportunity to experimentally test the predictions of the present work.

\section{Theoretical model}
\label{sec:model}
The system under consideration is an ultracold Fermi gas in which particles of opposite pseudo-spin interact via an $s$-wave contact potential. In the context of effective field theories, this system can be described in terms of a superfluid order parameter $\Psi(\mathbf{r},t)$, representing the bosonic field of Cooper pairs. Under the assumption that this field varies slowly in both space and time, a gradient expansion of the Euclidean-time action functional of the fermionic system can be performed, resulting in an effective action for the bosonic field:
\begin{align}
S[\Psi ]= \int_{0}^{\beta} d\tau \int d\mathbf{r} &\left[ \frac{D}{2}\left( \Psi^* \frac{\partial \Psi}{\partial \tau}- \frac{\partial \Psi^*}{\partial \tau }\Psi \right) +  \Omega_s + C  \left( \nabla_{\mathbf{r}} \Psi^* \cdot\nabla_{\mathbf{r}} \Psi \right) - E \left( \nabla_{\mathbf{r}} \vert \Psi \vert^2 \right)^2 \right. \notag \\
&+ \left. Q \frac{\partial \Psi^*}{\partial \tau} \frac{\partial \Psi}{\partial \tau} - R \left( \frac{\partial \vert \Psi \vert^2}{\partial \tau} \right)^2 \right] \label{eq:action}
\end{align}
where we use the natural units of  $\hbar = 1$, $2m = 1$, $E_F = 1$, and where $\beta$ is the inverse temperature. A more detailed explanation of this model can be found in Ref.\ \cite{THKTLDEpjB} and in Appendix \ref{sec:appEFT}, together with the analytical expressions for the thermodynamic potential $\Omega_s$ and the coefficients $C$, $D$, $E$, $Q$ and $R$ in terms of the average chemical potential $\mu$ and the interaction parameter $(k_F a_s)^{-1}$. The coefficients $C$, $E$, $Q$ and $R$ are only a function of the amplitude of the order parameter in bulk $\vert \Psi_{\infty} \vert$ (i.e.\ the superfluid gap). The coefficient $D$ and the thermodynamic potential $\Omega_s$, on the other hand, depend fully upon the local value of the amplitude of the order parameter \cite{THKTDPrA}. 
In this work, we assign to $\vert \Psi_{\infty} \vert$ and $\mu$ the mean-field values that are obtained by simultaneously solving the saddle-point gap and number equations \cite{THDevreeseTempere}. %At unitarity, this yields the values $\vert \Psi_{\infty} \vert = 0.69 \, E_F$, $\mu = 0.59 \, E_F$ and $T_c = 0.5 \, T_F$ for respectively the gap, chemical potential and critical temperature of the system. 
These background values could be further improved upon by, for example, including fluctuations around the saddle point, using the values of quantum Monte-Carlo simulations \cite{THCarlsonChang,THAstrakharchik,THCarlsonReddy} or even using the values derived from experimental measurements \cite{EXPKuSommer}.

From the Euclidian action functional \eqref{eq:action}, the real-time three-dimensional (3D) equation of motion for the pair field $\Psi(\mathbf{r},t)$ can be derived:
\begin{equation}
i \tilde{D}(\vert \Psi \vert^2) \frac{\partial \Psi}{\partial t} = -C \, \nabla_{\mathbf{r}}^2 \Psi + Q \frac{\partial^2 \Psi}{\partial t^2} + \left( \mathcal{A}(\vert \Psi \vert^2) + 2  E \, \nabla_{\mathbf{r}}^2 \vert \Psi \vert^2 - 2  R \frac{\partial^2 \vert \Psi \vert^2}{\partial t^2} \right) \Psi \label{eq:eqofmot}
\end{equation} 
where the coefficients $\tilde{D}$ and $\mathcal{A}$ are defined as
\begin{align}
 \tilde{D}
=\frac{\partial\left(  |\Psi|^2 D  \right)  }{\partial
\left(|\Psi|^2\right)}  \qquad \mathcal{A}  &  =\frac{\partial\Omega_{s} }{\partial \left( |\Psi|^2 \right)} \label{eq:A&D}
\end{align}
This equation is a type of non-linear Schr\"{o}dinger equation which is closely related to both the Gross-Pitaevskii equation for Bose-Einstein condensates \cite{THKTVPrA94} and the Ginzburg-Landau equation for Fermi superfluids \citep{THRanderiaSaDeMelo}. 
We find an analytical solution $\Psi_s(x)$ for a one-dimensional (1D) stationary dark soliton (also called a black soliton) by solving the time-independent equation
\begin{equation}
-C \, \partial_x^2 \Psi_s  +  \left( \mathcal{A}(\vert \Psi_s \vert^2) + 2 E \, \partial_x^2 \vert \Psi_s \vert^2  \right) \Psi_s = 0 \label{eq:stateqofmot}
\end{equation}
with boundary conditions
\begin{align}
&\lim_{x \rightarrow \pm \infty}\Psi(x) = \mp \Psi_{\infty} 
\end{align}
This solution for the order parameter possesses a phase jump $\pi$ and an amplitude equal to zero at the core of the soliton \cite{THKTDPrA,THLvAKTPrA}. %Because the pair density becomes zero in the center of a stationary soliton, it is often called a ``{black}'' soliton. 
%When studying the properties of solitonic excitations across the BEC-BCS crossover, it is often convenient to express the position variable $x$ in units of the healing length $\xi$ of the system. In these units, the width of solitonic excitations will be of the order of unity across the whole interaction domain. 
When studying solitonic excitations across the BEC-BCS crossover, it is convenient to express the length scale in units of the healing length $\xi$, which is here defined as the width of the soliton. An analytic expression for $\xi$ can be derived through a variational ansatz for the stationary soliton solution and a minimization of the EFT free energy (see also Appendix \ref{sec:appxi}). Using the system parameters of the experiment in Ref.\ \cite{EXPKu2}, the variational result for $\xi$ yields a soliton width of about 550 nanometers at unitarity.

The main assumption of the EFT model is that the order parameter $\Psi(\mathbf{r},t)$ varies slowly in both space and time \cite{THKTLDEpjB}. In terms of spatial fluctuations, this assumption corresponds to the condition that the pair field should vary over a spatial region larger than the pair correlation length (also referred to as the Pippard correlation length in the context of superconducting systems). In terms of the frequency and energy of the collective  excitations, the validity of the theory is mainly determined by the role of pair-breaking processes. Broken pairs are only present in the EFT in a local equilibrium state, there is no explicit pair-breaking dynamics. As a function of energy, pair-breaking processes become important at the bottom of the single-particle excitation spectrum ($2 \Delta$ in the BCS-regime, $2 \sqrt{\Delta^2 + \mu^2}$ in the BEC-regime), which means the validity of the EFT can only be guaranteed for fluctuations of which the energy lies below this threshold value. A detailed study of the validity of the model reveals that the theory is reliable, except in some cases in the BCS-regime \cite[Figure~5]{THLvAKTPrA}, where $\Delta$ becomes small and the pair correlation length becomes large. Earlier EFT calculations on the snake instability mechanism show a good agreement with results obtained from other theoretical formalisms \cite{THLvAKTSI} across the whole BEC-BCS crossover. This is a consequence of the fact that the unstable mode is in general a long-wavelength mode, with energies sufficiently far below $2 \Delta$. Accordingly, we also expect our current study on the nature of the instability mechanism to remain within the validity domain of the EFT.

\section{Results}
\label{sec:res}
\begin{figure}[b]
\centering
\centerline{
\includegraphics[scale=1.05]{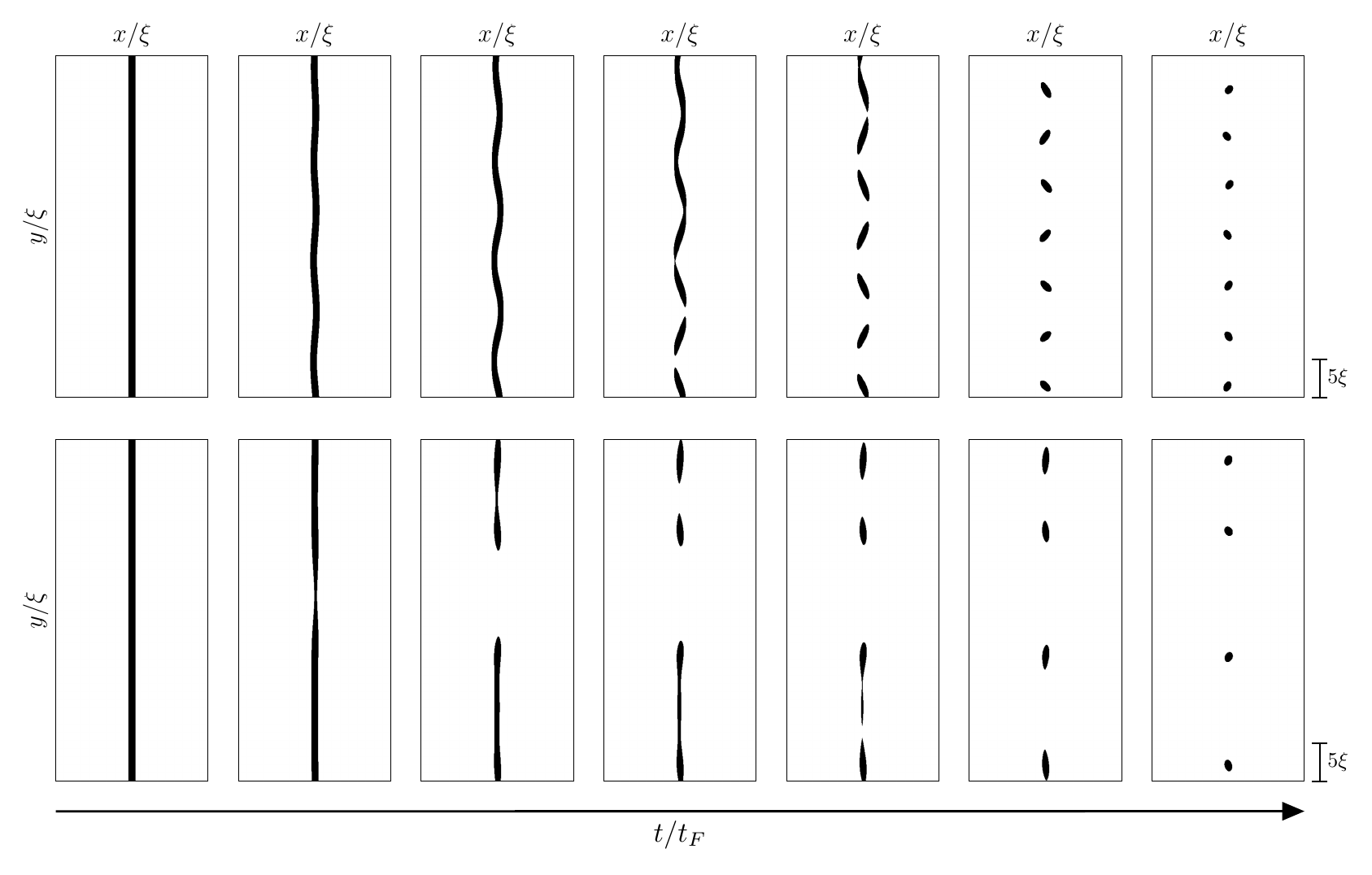}}
\caption{Evolution of the 2D pair density field $\vert \Psi \vert^2$ during the decay of a dark soliton for $(k_F a_s)^{-1} = 1$ (upper row) and $(k_F a_s)^{-1} = -2$ (lower row). The different columns correspond to snapshots at %$t/t_F = $ 0, 150, 160, 165, 168, 175 and 190 for the upper row, and $t/t_F = $ 0, 225, 236, 240, 244, 250 and 255 for the lower row. 
$t/t_F = $ 0, 150, 160, 165, 168, 175 and 190 for the upper row, and $t/t_F = $ 0, 225, 236, 240, 244, 250 and 270 for the lower row. 
Regions are black if $\vert \Psi \vert^2 < 0.1 \, \vert \Psi_{\infty} \vert^2$ and white otherwise. %The horizontal arrow underneath the frames indicates the direction of time.
}
\label{fig:evol}
\end{figure}
To investigate the dynamics of the soliton instability across the BEC-BCS crossover, we perform numerical simulations of the decay of a stationary dark soliton in a uniformly trapped quasi-two-dimensional (2D) superfluid Fermi gas, using the EFT's non-linear equation of motion \eqref{eq:eqofmot}. The initial state is constructed by extending the 1D stationary soliton solution $\Psi_s(x)$ into two dimensions (using periodic boundary conditions in the transverse direction) and adding a small amount of random noise to trigger the instability. The subsequent numerical time evolution is carried out by discretizing the space-time grid and applying a finite-difference fourth order Runge-Kutta (RK4) algorithm.  For the present calculations, the longitudinal and transverse system size are chosen to be respectively $L_x = 40 \, \xi$ and $L_y = 100 \, \xi$, while the spatial and temporal resolution are taken to be respectively $5\%$ of $\xi$ and $2\%$ of $t_F = \omega_F^{-1} = (E_F/\hbar)^{-1}$.  All calculations are carried out at a temperature $T = 0.001 \, T_F$, which for all practical purposes corresponds to zero temperature.  A more detailed explanation of this procedure is given in Appendix \ref{sec:appdis}.

Figure \ref{fig:evol} shows the evolution of the pair density $\vert \Psi(x,y) \vert^2$ of the superfluid during the decay of a stationary dark soliton. The upper row shows the evolution for $(k_F a_s)^{-1} = 1$ (BEC-regime), while the low row shows the evolution for $(k_F a_s)^{-1} = -2$ (BCS-regime). To make the dynamics of the soliton core more apparent, the spatial profile of the pair density is displayed using only two colors: black if $\vert \Psi(x,y) \vert^2 < 0.1 \, \vert \Psi_{\infty} \vert^2$ and white otherwise. In the BEC-regime, one can clearly observe the onset of snaking deformations of the depletion plane, which eventually break up the soliton into vortices. The manifestation of these snaking oscillations during the soliton decay is well-known from superfluid Bose gases. In the BCS-regime, on the other hand, the depletion plane does not exhibit snaking deformations during the decay process. Instead, the soliton core simply dissolves into vortices.%, which suggests the decay is driven by the creation of local currents across the soliton plane.

To further investigate the observed transition in the dynamics of the decay across the BEC-BCS crossover, we perform a perturbative study of the collective excitations around the stationary soliton solution. These excitations are represented by a small perturbation field $\delta \Psi(\mathbf{r},t) = \Psi(\mathbf{r},t) - \Psi_s(x)$.
%\begin{equation}
%\Psi(\mathbf{r},t) = \Psi_s(x) + \delta \Psi(\mathbf{r},t)
%\end{equation}
%Then, we plug this perturbed solution for the pair field into the equation of motion \eqref{eq:eqofmot} and expand the coefficients $\tilde{D}$ and $\mathcal{A}$ up to first order in the perturbation:
%\begin{align}
%\tilde{D}\big(\vert \Psi(\mathbf{r},t) \vert^2 \big) &= \tilde{D} \big(\vert \Psi_s(x) \vert^2 \big) +  \frac{\partial \tilde{D}}{\partial \vert \Psi \vert^2} \bigg\vert_{\vert \Psi_s \vert^2} \Psi_s(x) \Big[ \delta \Psi(\mathbf{r},t) + \delta \Psi^*(\mathbf{r},t) \Big] + ... \\
%%
%\mathcal{A}\big(\vert \Psi(\mathbf{r},t) \vert^2 \big) &= \mathcal{A} \big(\vert \Psi_s(x) \vert^2 \big) +  \frac{\partial \mathcal{A}}{\partial \vert \Psi \vert^2} \bigg\vert_{\vert \Psi_s \vert^2} \Psi_s(x) \Big[ \delta \Psi(\mathbf{r},t) + \delta \Psi^*(\mathbf{r},t) \Big] + ...
%\end{align}
%We will further use the notations
%\begin{equation}
%f_s = f \big(\vert \Psi_s(x) \vert^2 \big) \qquad \partial_s f_s = \frac{\partial f}{\partial \vert \Psi \vert^2} \bigg\vert_{\vert \Psi_s \vert^2} 
%\end{equation}
The equation of motion \eqref{eq:eqofmot} can be linearized with respect to this perturbation field in order to describe small fluctuations of the order parameter (see Appendix \ref{sec:applin}). To capture the physics of the instability more clearly, we also introduce the fields
\begin{equation}
P_{\pm}(\mathbf{r},t) = \left[ \delta \Psi(\mathbf{r},t) \pm \delta \Psi^*(\mathbf{r},t) \right] / 2 
\end{equation}
In the linearized theory, these fields correspond to fluctuations of respectively the amplitude and phase of the order parameter. Since we are interested in transverse unstable modes which propagate along the soliton plane in the $y$-direction and are localized around the soliton in the $x$-direction, we assume $P_{\pm}(\mathbf{r},t) = P_{\pm}(x) \, e^{i(k y - \omega t)}$. Then, the linearization of equation \eqref{eq:eqofmot} results in two coupled equations for the amplitude and phase field:
\begin{align}
\label{eq:eqamp}
&\alpha_1(x) \, P_+^{''} + \alpha_2(x) \, P_+^{'} + \Big( \alpha_3(x) - \omega^2 \, \alpha_4(x)   \Big)  \, P_+ = \, \omega \, \gamma(x) \, P_- \\
\label{eq:eqphase}
&\beta_1 \, P_-^{''} + \Big( \beta_2(x) - \omega^2 \, \beta_3 \Big) \, P_- =  \omega \, \gamma(x) \, P_+ 
\end{align}
The expressions for the position-dependent coefficients $\alpha_i(x)$, $\beta_i(x)$ and $\gamma(x)$ are given in Appendix \ref{sec:applin}.
In general, the two equations are coupled by a coupling coefficient $\gamma(x)$. However, this coupling coefficient becomes very small with respect to the other coefficients on the BCS-side of the interaction domain, causing the amplitude and phase modes to decouple for $(k_F a_s)^{-1} \ll -1$. On the BEC-side, on the other hand, $\gamma(x)$ becomes much larger, resulting in a strong coupling between the equations for $(k_F a_s)^{-1} \gg 1$. Consequently, tuning the interaction parameter $(k_F a_s)^{-1}$ from the BCS- to the BEC-regime increases the coupling between the amplitude and phase modes. 

The spectrum of eigenmodes $\omega(k)$ is obtained by numerically solving the coupled equations  \eqref{eq:eqamp} and \eqref{eq:eqphase}. In particular, the soliton will be unstable if there is at least one mode for which $\text{Im}(\omega) > 0$, since such a mode will grow exponentially in time. Additionally, we can determine for every mode whether it is more amplitude-like or phase-like in nature by calculating the norms of the eigenfields as
\begin{equation}
N_{\pm}  = \int \vert P_{\pm}(x) \vert^2 \, dx  
\end{equation}
and then using these quantities to define a mixing parameter
\begin{equation}
\eta = \frac{N_+}{N_+ + N_-}
\label{eq:eta}
\end{equation}
with $0 \leq \eta \leq 1$. For $\eta = 1$, the excitation is a pure amplitude mode, while for $\eta = 0$, it is purely a phase mode.

\begin{figure}[tb]
\centering
\centerline{
\includegraphics[scale=0.75]{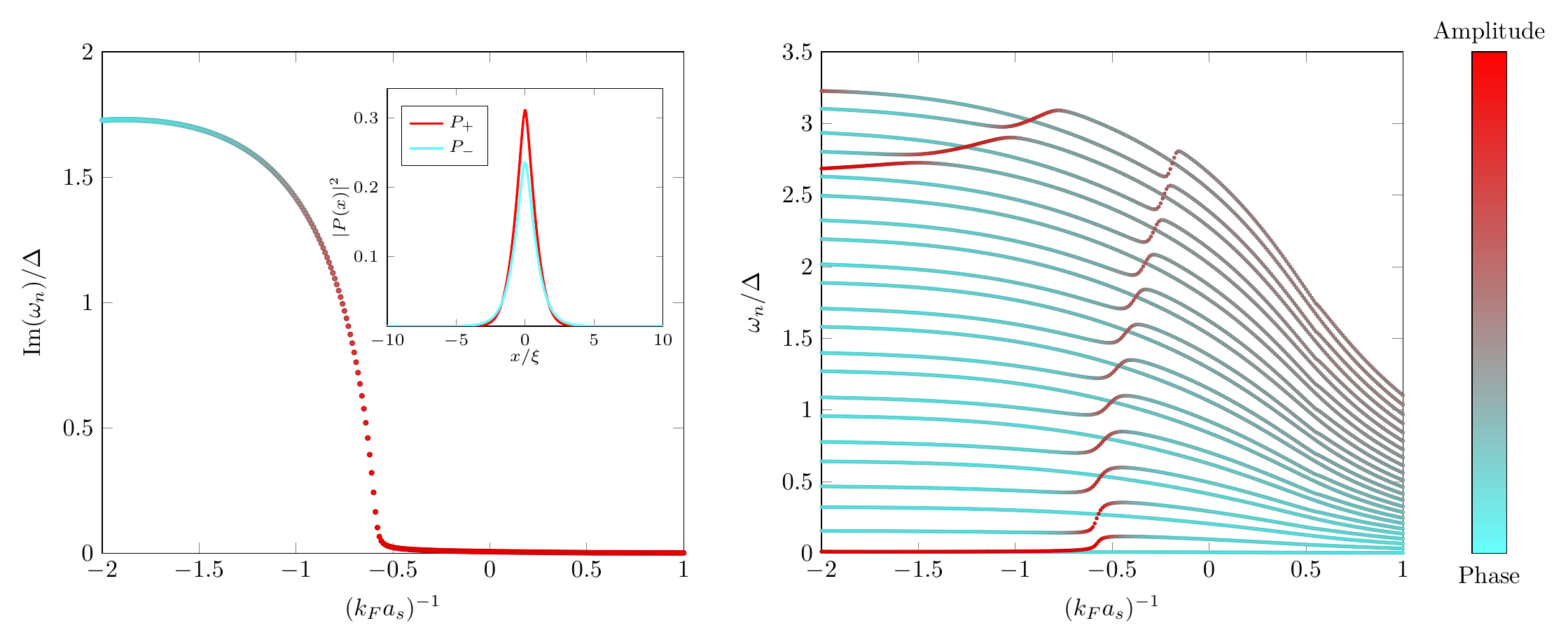}}
\caption{Eigenvalues of the collective modes in the presence of a stationary soliton for $k \approx 0$ in function of the interaction parameter $(k_F a_s)^{-1}$. The left figure shows the (positive) imaginary eigenvalues associated to the unstable mode, while the right figure shows the (positive) real eigenvalues. The color of each point indicates the nature of the associated eigenmode, based on the value of $\eta$ defined in equation \eqref{eq:eta}.  The inset of the left figure shows the spatial profiles of $P_+(x)$ and $P_-(x)$ for the unstable mode at $(k_F a_s)^{-1}=-0.9$.}
\label{fig:eigifoka}
\end{figure}

Figure \ref{fig:eigifoka} shows the real frequencies of several low-lying energy modes (right graph) and the imaginary part of the frequency of the unstable mode (left graph) in function of $(k_F a_s)^{-1}$ for $k \approx 0$.  The inset of the left graph shows an example of the profiles of the localized eigenfunctions $P_+(x)$ and $P_-(x)$ for the unstable mode at $(k_F a_s)^{-1} = -0.9$. Each point on the main graphs is assigned a color based on the value of $\eta$ for the associated eigenmode: as $\eta$ goes from 0 to 1, the assigned color shifts from blue to red. 
We observe that as we tune the interactions from the BCS-regime ( $(k_F a_s)^{-1} < 0$) to the BEC-regime ($(k_F a_s)^{-1} > 0$), the unstable mode changes from a phase-like mode to an amplitude-like mode. 
This shift in nature also appears to affect the behavior of the stable collective excitations across the interaction domain, inducing a steeply increasing amplitude mode which runs through the continuum of phase modes across the BEC-BCS crossover.
%At the BCS-side, the spectrum of stable modes initially contains a low-lying even amplitude mode. However, when the unstable mode changes nature and also becomes a low-lying amplitude mode, the stable amplitude mode is forced to transition into an even phase mode. As a result, all higher-lying even phase modes are shifted upwards and briefly become amplitude-like modes to cross the odd phase modes in between. All together, this creates the notion of a steeply increasing amplitude mode which runs through the continuum of phase modes across the crossover. 
%The same behavior in function of $(k_F a_s)^{-1}$ is observed at non-zero values of $k$.

\begin{figure}[tb]
\centering
\centerline{
\includegraphics[scale=0.6]{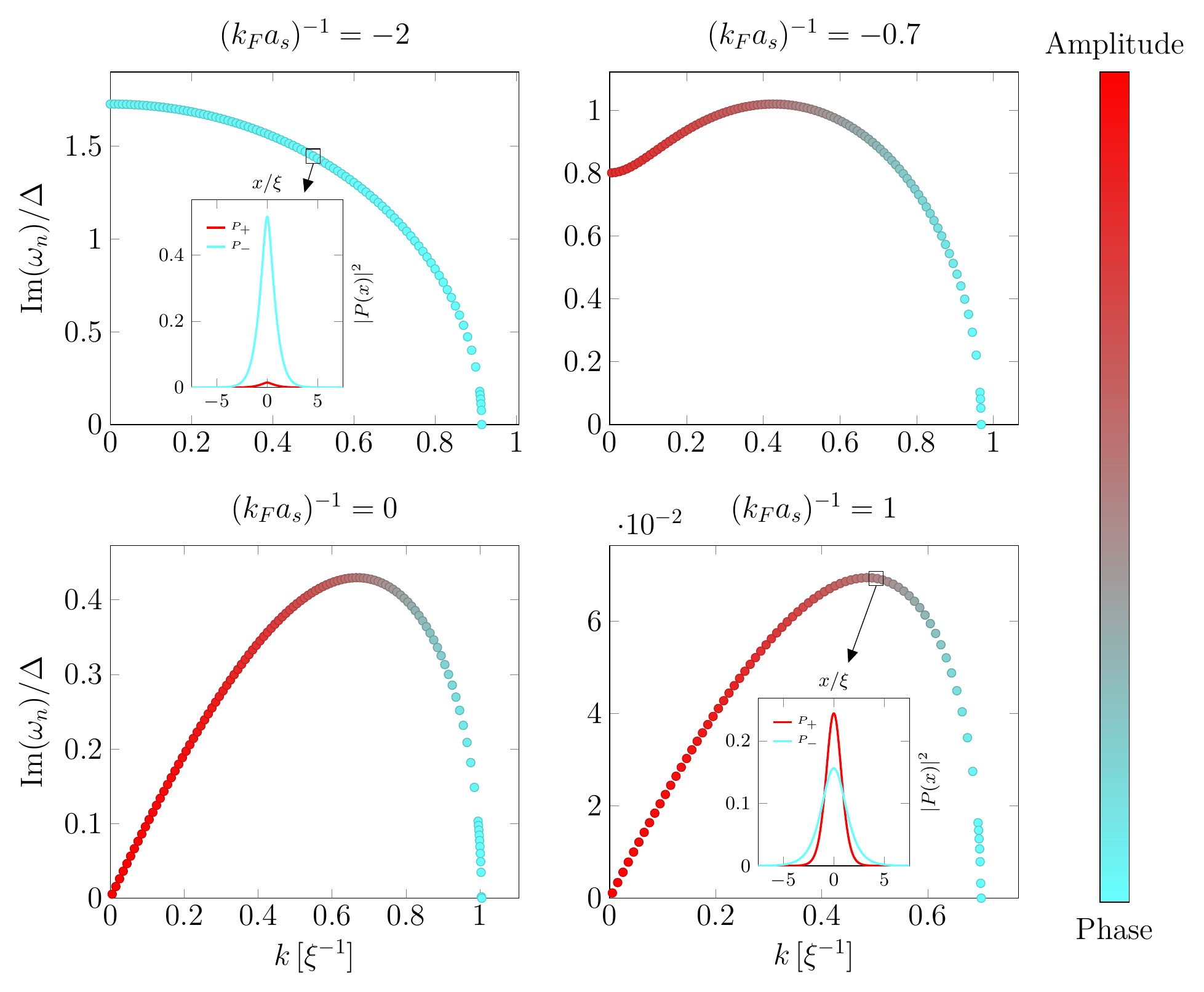}}
\caption{Imaginary part of the frequency of the unstable mode in function of $k$, for $(k_F a_s)^{-1} = -2$, $-0.7$, $0$ and $1$. The color of each point indicates the nature of the associated unstable mode, based on the value of $\eta$ defined in equation \eqref{eq:eta}. The insets show the localized eigenfunctions $P_+(x)$ and $P_-(x)$ for the modes
which have been observed to manifest during the decay processes in Figure \ref{fig:evol}.}
\label{fig:eigifok}
\end{figure}

Figure \ref{fig:eigifok} shows the imaginary part $\text{Im}(\omega)$ of the unstable modes for several values of $(k_F a_s)^{-1}$. The insets of the top left and bottom right graphs show the localized eigenfunctions $P_+(x)$ and $P_-(x)$ for %the typical modes which are expected 
the modes which have been observed to manifest during the decay processes in Figure \ref{fig:evol}.
It is clear that the unstable mode is a long-wavelength mode which only exists up to a maximum wave number $k_{c}$. We observe that, in the deep BCS-regime, the unstable mode is a phase mode across the whole $k$-domain, while closer to unitarity and in the BEC-regime, its nature strongly depends on the value of $k$: for lower values of $k$ it is amplitude-like, while for higher $k$ it is phase-like. %The transition between both fluctuation types seems to consistently occur at the maximum of the dispersion curve. 
We can interpret these observations in terms of the width of the system, as for a given transverse system size $L_y$, the largest transverse mode which fits into the system corresponds to the wave number $k = 2 \pi/L_y$. If it is assumed that the mode with the largest growth rate will be the one to manifest itself, one would need a considerably small transverse width $L_y$ at unitarity or in the BEC-regime to observe the phase-like modes close to $k_{c}$. In the deep BCS-regime, on the other hand, the nature of the unstable mode will always be phase-like, independent of the system width. Considering the typical system sizes which are currently employed in dark soliton experiments in superfluid Fermi gases \cite{EXPYefsah,EXPKu,EXPKu2}, we expect the phase-like character of the unstable mode will only be observable in the BCS-regime.

Combining the results of the numerical simulations and the perturbative analysis now allows us to explain the change in the behavior of the soliton instability observed in Figure \ref{fig:evol}. In the BEC-regime, the mode with the largest growth rate has a finite wave number and is a mix between an amplitude-like and a phase-like mode (right bottom graph of Figure \ref{fig:eigifok}). The amplitude-like contribution causes the characteristic snaking oscillations of the soliton core \footnote{The $k=0$ amplitude mode $\delta \Psi \propto \delta x \frac{d \Psi_s}{d x}$ corresponds to a uniform shift of the soliton in the $x$-direction over a distance $\delta x$. For finite $k$, the magnitude and direction of this translation vary along the soliton plane, leading to the creation of the characteristic snaking pattern.}, while the phase-like contribution creates local Josephson currents, leading to the manifestation of vortices. In the deep BCS-regime, on the other hand, unstable modes are purely phase-like (top left graph of Figure \ref{fig:eigifok}). As such, they cause vortex nucleation without the occurrence of snaking deformations. Therefore, we suggest that, on the BCS-side of the interaction domain, ``{Josephson instability}'' might be a more suitable name than snake instability to describe the unstable mode, as no observable snaking pattern is present. Moreover, since in the deep BCS-regime the imaginary part of the frequency reaches its maximum value at $k=0$, the number of generated vortices is expected to be smaller than in the BEC-regime, as is indeed the case in Figure \ref{fig:evol}.

%Combining the observations of the numerical simulations and the perturbative analysis demonstrates that the amplitude-to-phase crossover of the unstable mode has a profound impact on the dynamics of the decay. On the BEC-side of the interaction domain, the dark soliton is destroyed as a result of the characteristic snaking deformations of the depletion plane, which are induced by the amplitude-like aspect of the fluctuations. On the BCS-side, the phase-like fluctuations instead create local Josephson currents across the soliton plane, which break up the core without distinctly changing its position. Therefore, we suggest that, on the BCS-side of the interaction domain, ``{Josephson instability}'' might be a more suitable name than snake instability to describe the soliton decay, since no observable snaking deformations are present in this regime.
%

\section{Conclusions}
\label{sec:concl}
The combined observations of the nonlinear dynamics and the perturbative analysis of the soliton decay indicate that the crossover in the behavior of the soliton instability is caused by the fact that the nature of the unstable mode changes from amplitude-like to phase-like when one moves from the BEC- to the BCS-side of the interaction domain. 
In the BCS-regime, where the unstable mode is purely phase-like in nature, the creation of local Josephson currents leads to the disintegration of the soliton core into vortices without changing the soliton's position. In the BEC-regime, on the other hand, where the nature of the unstable mode is a mix between amplitude-like and phase-like, the nucleation of vortices is preceded by the onset of snaking deformations of the soliton core. We proposed to name the soliton instability in the BCS-regime the ``{Josephson instability}'', in order to distinguish it from the snake instability in the BEC-regime.\\
We can compare our predictions with the results of the experiment carried out in Ref.\ \cite{EXPKu2}, in which the authors imaged the decay of a dark soliton in a 3D harmonically trapped superfluid Fermi gas at unitarity. By applying the system parameters of this experiment to the case of a uniformly trapped superfluid, we obtain that $t_F \approx 5.2 \times 10^{-5}$ seconds, which yields a typical decay time on the order of (tens of) milliseconds for our numerical simulations across the BEC-BCS crossover. This time scale is in good agreement with the decay time observed in Ref.\ \cite{EXPKu2}, even though the experiment was performed in a harmonically trapped system. We furthermore observe that, on some of the experimental images in Ref.\ \cite{EXPKu2}, the decay process actually looks like a crossover between the snake instability mechanism and the aforementioned Josephson instability mechanism, indicating that, in practice, the predicted transition in the dynamics of the decay might already be observable around unitarity. This hypothesis will have to be investigated more thoroughly in future experiments. We suspect that large, quasi-2D systems, like the one employed in the experiment in Ref.\ \cite{EXPPark}, might be most suitable to test the current predictions, as they closely resemble the theoretical system described in this work. %Finally, it is also important to note that, while the images in Figure \ref{fig:evol} show the evolution of the pair field density, actual experiments image the evolution of the fermion density. At sufficiently low temperatures, however, the two quantities should be related closely enough to test and observe the predicted dynamics.

%The calculations carried out in this work assume the background of the superfluid to be uniform. While it is true that ultracold gases are traditionally studied in harmonic confinements, the recent realization of box-like optical traps \cite{EXPGauntSchmidutz} can provide the opportunity to test the predictions of this work in experiment. 

\acknowledgments W.\ Van Alphen acknowledges financial support in the form of a Ph.\ D.\ fellowship of the Research Foundation - Flanders (FWO). This research was supported by the University Research Fund (BOF) of the University of Antwerp, by the Flemish Research Foundation (FWO-Vl) project nr G.0429.15.N., by JSPS KAKENHI Grants No. JP17K05549 and JP17H02938, and in part by the Osaka City University (OCU) Strategic Research Grant 2017 for young researchers.

\appendix
\section{Overview of the EFT}
\label{sec:appEFT}
\def\theequation{A.\arabic{equation}}
\setcounter{equation}{0}
In this section we provide a brief overview of the EFT model and the expressions for the EFT expansion coefficients. More detailed derivations and explanations can be found in Ref.\ \cite{THDevreeseTempere, THKTLDEpjB, THLombardiPhD}. \\
The system of interest is an ultracold Fermi gas, in which particles of opposite pseudo-spin interact via an $s$-wave contact potential. The Euclidian-time action functional of this system can be written down in terms of the fermionic (Grassmann) fields $\psi_{\sigma}(\mathbf{x},\tau)$ and $\bar{\psi}_{\sigma}(\mathbf{x},\tau)$:
\begin{multline}
S[\psi] = \int_0^\beta d\tau \int d\textbf{x} \left[\sum_{\sigma\in\lbrace \uparrow,\downarrow\rbrace}\bar{\psi}_\sigma(\textbf{x},\tau)\left(\frac{\partial}{\partial \tau} -\nabla^2_{\textbf{x}}-\mu_\sigma\right)\psi_\sigma(\textbf{x},\tau)  +g \, \bar{\psi}_\uparrow(\textbf{x},\tau)\bar{\psi}_\downarrow(\textbf{x},\tau)\psi_\downarrow(\textbf{x},\tau)\psi_\uparrow(\textbf{x},\tau)\right]
\label{eq:FermionicAction}
\end{multline}
where $g$ is the strength of the contact interaction and the label $\sigma$ denotes the spin degree of freedom. The quartic interaction term can be decoupled through the Hubbard-Stratonovich (HS) transformation, which introduces the bosonic pair field $\Psi(\mathbf{x},\tau)$ (the HS field is often also denoted as $\Delta$, but here we use $\Psi$ to emphasize its interpretation as a position- and time-dependent order parameter for the system) \cite{THDevreeseTempere}. The fermionic degrees of freedom can then be integrated out. If we assume that the pair field $\Psi(\mathbf{x},\tau)$ only varies slowly around its constant background value $\Psi_{\infty}$, we can perform a gradient expansion around $\Psi_{\infty}$ up to second order in the spatial and temporal derivatives of $\Psi(\mathbf{x},\tau)$ \cite{THKTLDEpjB}. This results in the following Euclidian-time effective action functional for the bosonic pair field:
\begin{align}
S_{\text{EFT}}[\Psi ]= \int_{0}^{\beta} d\tau \int d\mathbf{r} &\left[ \frac{D}{2}\left( \bar{\Psi}\frac{\partial \Psi}{\partial \tau}- \frac{\partial \bar{\Psi}}{\partial \tau }\Psi \right) +  \Omega_s + C  \left( \nabla_{\mathbf{r}} \bar{\Psi} \cdot\nabla_{\mathbf{r}} \Psi \right) - E \left( \nabla_{\mathbf{r}} \vert \Psi \vert^2 \right)^2 \right. \notag \\
&+ \left. Q \frac{\partial \bar{\Psi}}{\partial \tau} \frac{\partial \Psi}{\partial \tau} - R \left( \frac{\partial \vert \Psi \vert^2}{\partial \tau} \right)^2 \right] \label{eq:actionapp}
\end{align}
This effective action functional forms the starting point \eqref{eq:action} for our study of the snake instability in the main work. The thermodynamic potential $\Omega_s$ is given by:
\begin{align}
\label{eq:omega}
\Omega_s &= -\frac{1}{8 \pi k_F a_s}\vert \Psi \vert^2 - \int \frac{d\mathbf{k}}{(2 \pi)^3} \left\lbrace \frac{1}{\beta} \ln[2 \cosh(\beta E_{\mathbf{k}}) + 2 \cosh(\beta \zeta)] - \xi_{\mathbf{k}} - \frac{\vert \Psi \vert^2}{2 k^2}  \right\rbrace 
\end{align}
while the gradient expansion coefficients $D$, $C$, $E$, $Q$ and $R$ are defined as
\begin{align}
D &= \int \frac{d \mathbf{k}}{(2 \pi)^3} \frac{\xi_{\mathbf{k}}}{\vert \Psi \vert^2} [ f_1(\beta,\xi_{\mathbf{k}},\zeta) - f_1(\beta,E_{\mathbf{k}},\zeta) ]  \label{eq:d} \\
C &= \int \frac{d \mathbf{k}}{(2 \pi)^3} \frac{k^2}{3 m} f_2 (\beta,E_{\mathbf{k}},\zeta) \label{eq:c} \\
E &= 2 \int \frac{d \mathbf{k}}{(2 \pi)^3} \frac{k^2}{3 m} \, \xi_{\mathbf{k}}^2 \, f_4(\beta,E_{\mathbf{k}},\zeta)  \label{eq:e} \\
Q &= \frac{1}{2 \vert \Psi \vert^2} \int \frac{d \mathbf{k}}{(2 \pi)^3} [f_1(\beta,E_{\mathbf{k}},\zeta) -(E_{\mathbf{k}}^2 + \xi_{\mathbf{k}}^2) f_2(\beta,E_{\mathbf{k}},\zeta)]  \label{eq:q} \\
R &= \frac{1}{2 \vert \Psi \vert^2} \int \frac{d\mathbf{k}}{(2 \pi)^3} \left[ \frac{ f_1(\beta,E_{\mathbf{k}},\zeta) + (E_{\mathbf{k}}^2 - 3 \xi_{\mathbf{k}}^2) f_2(\beta,E_{\mathbf{k}},\zeta) }{3 \vert \Psi \vert^2} \right. \notag \\*
& \hspace{7em} + \left. \frac{4(\xi_{\mathbf{k}}^2 - 2 E_{\mathbf{k}}^2)}{3} f_3(\beta,E_{\mathbf{k}},\zeta) + 2 E_{\mathbf{k}}^2 \vert \Psi \vert^2 f_4(\beta,E_{\mathbf{k}},\zeta) \right]  \label{eq:r}
\end{align}
The functions $f_j(\beta,\epsilon,\zeta)$ in the above expressions are defined by
\begin{align}
f_{j}(\beta,\epsilon,\zeta)=\frac{1}{\beta}\sum_{n}\frac{1}{\left[\left(\omega_n-i\zeta\right)^2+\epsilon^2\right]^j}
\end{align}
with the fermionic Matsubara frequencies $\omega_n=(2n+1)\pi/\beta$.
In this treatment, the chemical potentials of the two pseudo-spin species $\mu_{\uparrow}$ and $\mu_{\downarrow}$ are combined into the average chemical potential $\mu = (\mu_{\uparrow} + \mu_{\downarrow})/2$ and the  imbalance chemical potential $\zeta = (\mu_{\uparrow} - \mu_{\downarrow})/2$, the latter determining the difference between the number of particles in each spin-population. The quantity $\xi_{\mathbf{k}} = \frac{k^2}{2m} - \mu$ is the dispersion relation for a free fermion, $E_{\mathbf{k}} = (\xi_{\mathbf{k}}^2 + \vert \Psi_{\mathbf{x},\tau} \vert^2)^{1/2}$ is the local Bogoliubov excitation energy, and $a_s$ is the $s$-wave scattering length that determines the strength and the sign of the contact interaction. In absence of spatial and temporal variations, the thermodynamic potential $\Omega_s$ determines the value of the pair-breaking gap $\vert \Psi_{\infty} \vert$ for the uniform system through the saddle-point gap equation
\begin{equation}
\frac{\partial \Omega_s}{\partial \vert \Psi \vert^2} \Psi = 0
\label{eq:gap}
\end{equation}
This equation is solved self-consistently together with the number equation to obtain the correct values of $\vert \Psi_{\infty} \vert$ and $\mu$ for a given set of system parameters.\\
In principle, all expansion coefficients \eqref{eq:omega}--\eqref{eq:r} fully depend upon the order parameter $\Psi(\mathbf{x},\tau)$, but in practice, we assume that the coefficients associated with the second order derivatives of the pair field can be kept constant and equal to their bulk value, since retaining their full space-time dependence would lead us beyond the second-order approximation of the gradient expansion. This means that in expressions \eqref{eq:c}, \eqref{eq:e}, \eqref{eq:q} and \eqref{eq:r} for the coefficients $C$, $E$, $Q$ and $R$, we set $\vert \Psi(\mathbf{x},\tau) \vert^2 \rightarrow \vert \Psi_{\infty} \vert^2$ and $E_{\mathbf{k}} \rightarrow E_{\mathbf{k},\infty} =  (\xi_{\mathbf{k}}^2 + \vert \Psi_{\infty} \vert^2)^{1/2}$. For the thermodynamic potential $\Omega_s$ and the coefficient $D$, on the other hand, the full space-time dependence of the order parameter is preserved.

From the Euclidian-time action functional \eqref{eq:actionapp}, the EFT equation of motion for the pair field $\Psi(\mathbf{r},t)$ is found to be
\begin{equation}
i \tilde{D}(\vert \Psi \vert^2) \frac{\partial \Psi}{\partial t} = -C \, \nabla_{\mathbf{r}}^2 \Psi + Q \frac{\partial^2 \Psi}{\partial t^2} + \left( \mathcal{A}(\vert \Psi \vert^2) + 2  E \, \nabla_{\mathbf{r}}^2 \vert \Psi \vert^2 - 2  R \frac{\partial^2 \vert \Psi \vert^2}{\partial t^2} \right) \Psi \label{eq:eqofmotapp}
\end{equation} 
where the coefficients $\tilde{D}$ and $\mathcal{A}$ are defined as
\begin{align}
 \tilde{D}
=\frac{\partial\left(  |\Psi|^2 D  \right)  }{\partial
\left(|\Psi|^2\right)}  \qquad \mathcal{A}  &  =\frac{\partial\Omega_{s} }{\partial \left( |\Psi|^2 \right)} \label{eq:A&Dapp}
\end{align}
The first term on the right-hand side of the equation can be identified as a kinetic energy term, while the non-linear term represents a system-inherent potential for the field. The ratio $\tilde{D}/C$ can be interpreted as a renormalization factor for the mass of the fermion pairs \cite{THKTVPrA94} and the coefficient $\mathcal{A}$ determines the uniform background value of the system, since $\mathcal{A}(\Psi) \, \Psi = 0$ is nothing but the aforementioned gap equation \eqref{eq:gap}. It has been verified that in the deep BEC-limit $\left( 1/k_F a_S \gg 1 \right)$, the equation correctly tends to the Gross-Pitaevskii equation for bosons with a mass $M=2m$ and an s-wave boson-boson scattering length $a_B = 2 \, a_s$ \citep{THLombardiPhD}. 

\section{Variational derivation of the healing length}
\label{sec:appxi}

\begin{figure}[htbp]
\centering
\centerline{
\includegraphics[scale=0.8]{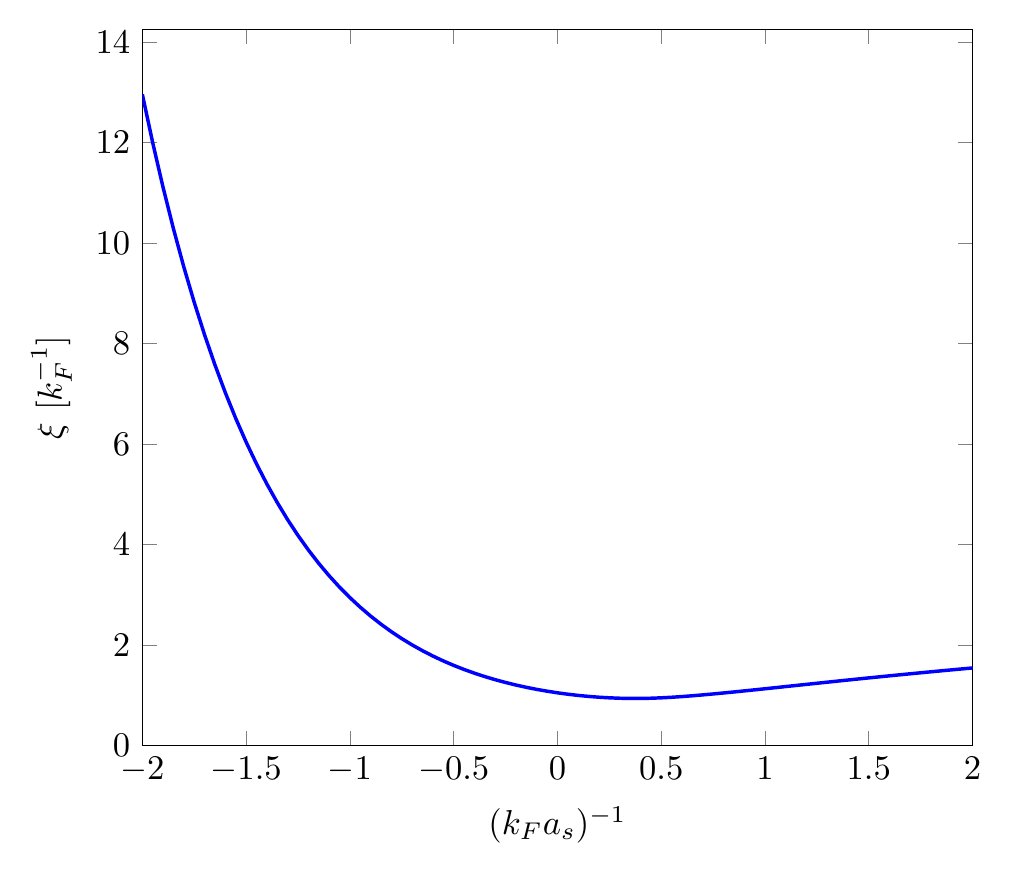}}
\caption{Variational estimate of the healing length in function of the interaction parameter.}
\label{fig:xivar}
\end{figure}

We can derive an analytic expression for the healing length $\xi$ associated with a stationary (black) soliton in a superfluid by considering a variational ansatz for the wavefunction and minimizing the free energy of the system. Since the wavefunction of a stationary soliton is expected to be real and anti-symmetric, we assume a hyperbolic tangent model for the pair field:
\begin{equation}
\Psi(x) = \Delta \, \tanh\left( \frac{x}{\sqrt{2} \, \xi} \right)
\label{eq:wavefunction}
\end{equation}
The EFT free energy for a one-dimensional Fermi superfluid is given by
\begin{align}
F_{\text{EFT}}[\Psi] = \int_{-\infty}^{\infty} dx \left[ X(\vert \Psi \vert^2) + C \, \partial_x \bar{\Psi} \, \partial_x \Psi - E \left( \partial_x \vert \Psi \vert^2 \right)^2 \right]
\label{eq:Fgeneral}
\end{align}
with
\begin{equation}
X(\vert \Psi \vert^2) = \Omega_{s}(\vert \Psi \vert^2) - \Omega_{s}(\vert \Psi_{\infty} \vert^2)
\end{equation}
The subtraction of the term $\Omega_{s}(\vert \Psi_{\infty} \vert^2)$ indicates that the energy is measured with respect to the energy of the uniform system.  By substituting the ansatz \eqref{eq:wavefunction} for the pair field into the free energy and making a change of integration variable $u = x / (\sqrt{2} \xi)$, we obtain
\begin{equation}
F = \sqrt{2} \, \int_{-\infty}^{\infty} du \left[ \xi \, X(u)  + \frac{\tilde{C} \, \Delta^2}{2 \, \xi} \, \text{sech}^4\left(u\right) - \frac{\tilde{E} \, \Delta^4}{\xi} \, \text{sech}^4\left( u \right) \, \tanh^2\left( u \right) \right]
\end{equation}
The integrals in the second and third term can be computed analytically.  The derivative of $F$ with respect to the variational parameter $\xi$ then becomes:
\begin{equation}
\frac{dF}{d\xi} = \sqrt{2} \left[ \, \int_{-\infty}^{\infty} X(u) \, du - \frac{2 \, \tilde{C} \, \Delta^2}{3 \, \xi^2} + \frac{4 \, \tilde{E} \, \Delta^4}{15 \, \xi^2} \right]
\end{equation}
By setting the above equation equal to zero, we find the following variational expression for the healing length:
\begin{equation}
\xi = \sqrt{\frac{10 \, \tilde{C} \, \Delta^2 - 4 \, \tilde{E} \, \Delta^4}{15 \, B}}
\end{equation}
with
\begin{equation}
B = \int_{-\infty}^{\infty} X(u) \, du
\end{equation}
Figure \ref{fig:xivar} shows the behavior of this quantity in function of the interaction parameter $(k_F a_s)^{-1}$. A similar expression was derived in the context of the EFT for the width of a vortex core in Ref.\ \cite{THVerhelstPhysC}. A more extensive study on the healing length of a fermionic superfluid across the BEC-BCS crossover can be found in Ref.\ \cite{THPalestiniStrinati}.

\section{Discretization and evolution of the equation of motion}
\label{sec:appdis}

%\subsection{Finite-difference algorithm}
In this section we elaborate on how the EFT equation of motion \eqref{eq:eqofmot} is discretized and solved numerically using the explicit RK4 algorithm.
We introduce a field $\phi(\mathbf{r},t)$ such that
\begin{equation}
\phi = \frac{\partial \Psi}{\partial t} 
\label{eq:dt1}
\end{equation}
and
\begin{equation}
\bar{\phi} = \overline{\frac{\partial \Psi}{\partial t}} = \frac{\partial \bar{\Psi}}{\partial t}
\end{equation}
Substituting this into the equation of motion and making use of the fact that
\begin{equation}
\frac{\partial^2 \vert \Psi \vert^2}{\partial t^2} = \bar{\Psi} \frac{\partial^2 \Psi}{\partial t^2} + 2 \frac{\partial \bar{\Psi}}{\partial t} \frac{\partial \Psi}{\partial t} + \Psi \frac{\partial^2 \bar{\Psi}}{\partial t^2}
\end{equation}
we have
\begin{equation}
i \tilde{D}(\vert \Psi \vert^2) \phi = -\tilde{C} \, \nabla_{\mathbf{r}}^2 \Psi + Q \frac{\partial \phi}{\partial t} + \left( \mathcal{A}(\vert \Psi \vert^2) + \tilde{E} \, \nabla_{\mathbf{r}}^2 \vert \Psi \vert^2 - \tilde{R} \left( \bar{\Psi} \frac{\partial \phi}{\partial t} + 2 \vert \phi \vert^2 + \Psi \frac{\partial \bar{\phi}}{\partial t} \right) \right) \Psi
\label{eq:dtphidtphib} 
\end{equation}
In order to get an equation of the form $\partial_t \phi = ... \,$, we take the complex conjugate of \eqref{eq:dtphidtphib}, find an expression for $\partial_t \bar{\phi}$ in function of $\partial_t \phi$ and substitute this back into \eqref{eq:dtphidtphib}, yielding
\begin{align}
\frac{\partial \Phi}{\partial t} &= \frac{1}{Q \, (Q - 2 \tilde{R} \vert \Psi \vert^2)}  \left[ -Q \left( \mathcal{A} + \tilde{E} \, \nabla_{\mathbf{r}}^2 \vert \Psi \vert^2 - 2 \tilde{R} \vert \phi \vert^2  \right) \Psi + i \tilde{D} \left( Q \phi - \tilde{R} \, \Psi \left( \bar{\phi} \Psi + \phi \bar{\Psi} \right) \right) \right. \notag \\
 &+ \left. \tilde{C} \left( \Psi^2 \tilde{R} \, \nabla_{\mathbf{r}}^2 \bar{\Psi}   + \nabla_{\mathbf{r}}^2 \Psi (Q - \tilde{R} \vert \Psi \vert^2)  \right) \right] 
 \label{eq:dt2}
\end{align}
Equations \eqref{eq:dt1} and \eqref{eq:dt2} form a system of two coupled partial differential equations of the form:
\begin{align}
\frac{\partial \Psi}{\partial t} &= f(\phi) \\
\frac{\partial \phi}{\partial t} &= g(\Psi,\phi)
\end{align}
where $f(\phi) = \phi$ and $g(\Psi,\phi)$ is given by \eqref{eq:dt2}. In the case of a 2D system, we use finite mesh widths $\Delta x$ and $\Delta y$ and a finite time step $\Delta t$ to discretize space-time into a grid of $L \times M \times N$ points by writing $x_l = l \Delta x$ with $l = 1,...,L$, $y_m = m \Delta y$ with $m = 1,...,M$ and $t_n = n \Delta t $ with $n = 1,...,N$. This allows us to approximate the spatial derivatives by central finite difference formulas:
\begin{align}
\frac{\partial^2 \Psi(x,y,t)}{\partial x^2} = \frac{\Psi_{l+1,m,n} - 2 \, \Psi_{l,m,n} + \Psi_{l-1,m,n}}{\Delta x^2} \\[10pt]
\frac{\partial^2 \Psi(x,y,t)}{\partial y^2} = \frac{\Psi_{l,m+1,n} - 2 \, \Psi_{l,m,n} + \Psi_{l,m-1,n}}{\Delta y^2}
\end{align}
where we use the notation $\Psi_{l,m,n} = \Psi(x_l,y_m,t_n)$. Since we expect the superfluid to assume its uniform bulk value sufficiently far from the soliton, we require the derivatives of he fields to be zero at the $x$-boundaries of the grid. In the $y$-direction, we apply periodic boundary conditions. If we now know the values $\Psi_{l,m,n}$ and $\phi_{l,m,n}$ at a certain time step $t_n$ for all positions $x_l$ and $y_m$, the explicit RK4 method allows us to calculate for every position the values $\Psi_{l,m,n+1}$ and $\phi_{l,m,n+1}$ of the next time step by using the following algorithm  \citep{THSuliMayers}:
\begin{align}
&p_{1_{l,m,n}} = f(\phi_{l,m,n}) \\
&p_{2_{l,m,n}} = g(\Psi_{l,m,n},\phi_{l,m,n}) \\
&q_{1_{l,m,n}} = f(\phi_{l,m,n}+p_{2_{l,m,n}}/2) \\
&q_{2_{l,m,n}} = g(\Psi_{l,m,n}+p_{1_{l,m,n}}/2,\phi_{l,m,n}+p_{2_{l,m,n}}/2) \\
&r_{1_{l,m,n}} = f(\phi_{l,m,n}+q_{2_{l,m,n}}/2) \\
&r_{2_{l,m,n}} = g(\Psi_{l,m,n}+q_{1_{l,m,n}}/2,\phi_{l,m,n}+q_{2_{l,m,n}}/2) \\
&s_{1_{l,m,n}} = f(\phi_{l,m,n}+r_{2_{l,m,n}}) \\
&s_{2_{l,m,n}} = g(\Psi_{l,m,n}+r_{1_{l,m,n}},\phi_{l,m,n}+r_{2_{l,m,n}}) \\
&\Psi_{l,m,n+1} = \Psi_{l,m,n} + \frac{\Delta t}{6}(p_{1_{l,m,n}} + 2 \, q_{1_{l,m,n}} + 2 \, r_{1_{l,m,n}} + s_{1_{l,m,n}}) \\
&\phi_{l,m,n+1} = \phi_{l,m,n} + \frac{\Delta t}{6}(p_{2_{l,m,n}} + 2 \, q_{2_{l,m,n}} + 2 \, r_{2_{l,m,n}} + s_{2_{l,m,n}})
\end{align}
This scheme can be repeated until the solution has been evolved up to the desired point in time.

\section{Linearization of the equation of motion}
\label{sec:applin}

To describe small fluctuations of the pair field, we add a  perturbation field $\delta \Psi(\mathbf{r},t)$ to the stable soliton solution $\Psi_s(x)$:
\begin{equation}
\Psi(\mathbf{r},t) = \Psi_s(x) + \delta \Psi(\mathbf{r},t)
\end{equation}
We plug this perturbed solution into the equation of motion \eqref{eq:eqofmot} and expand the coefficients $\tilde{D}$ and $\mathcal{A}$ (which depend on the local value of the order parameter) up to first order in the perturbation field:
\begin{align}
\tilde{D}\big(\vert \Psi(\mathbf{r},t) \vert^2 \big) &= \tilde{D} \big(\vert \Psi_s(x) \vert^2 \big) +  \frac{\partial \tilde{D}}{\partial \vert \Psi \vert^2} \bigg\vert_{\vert \Psi_s \vert^2} \Psi_s(x) \Big[ \delta \Psi(\mathbf{r},t) + \delta \Psi^*(\mathbf{r},t) \Big] + ... \\
\mathcal{A}\big(\vert \Psi(\mathbf{r},t) \vert^2 \big) &= \mathcal{A} \big(\vert \Psi_s(x) \vert^2 \big) +  \frac{\partial \mathcal{A}}{\partial \vert \Psi \vert^2} \bigg\vert_{\vert \Psi_s \vert^2} \Psi_s(x) \Big[ \delta \Psi(\mathbf{r},t) + \delta \Psi^*(\mathbf{r},t) \Big] + ...
\end{align}
In order to study explicitly the character of amplitude and phase modes, we introduce the fields
\begin{equation}
P_{\pm}(\mathbf{r},t) = \left[ \delta \Psi(\mathbf{r},t) \pm \delta \Psi^*(\mathbf{r},t) \right] / 2 
\end{equation}
If we then collect all terms of first order in the fluctuation fields, we find two linearized equations
\begin{align}
\label{eq:eqampapp}
&\alpha_1(x) \, P_+^{''} + \alpha_2(x) \, P_+^{'} + \Big( \alpha_3(x) - \omega^2 \, \alpha_4(x)   \Big)  \, P_+ = \, \omega \, \gamma(x) \, P_- \\
\label{eq:eqphaseapp}
&\beta_1 \, P_-^{''} + \Big( \beta_2(x) - \omega^2 \, \beta_3 \Big) \, P_- =  \omega \, \gamma(x) \, P_+ 
\end{align}
where the coefficients $\alpha_i(x)$, $\beta_i(x)$ and $\gamma(x)$ are given by
\begin{align}
&\alpha_1(x) = -\Big(\tilde{C} - 2 \, \tilde{E} \, \Psi_s^2(x)\Big) \\
&\alpha_2(x) = 4 \, \tilde{E} \, \Psi_s(x) \, \partial_x \Psi_s(x) \\
&\alpha_3(x) = \tilde{C} \, k^2 + \mathcal{A}_s(x) + 2 \, \partial_s \mathcal{A}_s(x) \, \Psi_s^2(x) + 2 \, \tilde{E} \, \Big( \partial_x \Psi_s(x) \Big)^2 + 4 \, \tilde{E} \, \Psi_s(x) \, \partial_x^2 \Psi_s(x) \\
&\alpha_4(x) = Q - 2 \, \tilde{R} \, \Psi_s^2(x) \\
&\beta_1 = -\tilde{C} \\
&\beta_2(x) = \tilde{C} \, k^2 +  \mathcal{A}_s(x) + 2 \, \tilde{E} \, \Big( \partial_x \Psi_s(x) \Big)^2 + 2 \, \tilde{E} \, \Psi_s(x) \, \partial_x^2 \Psi_s(x) \\
&\beta_3 = Q \\
&\gamma(x) = \tilde{D}_s(x)
\end{align}
Here, we have used the notations
\begin{equation}
f_s = f \big(\vert \Psi_s(x) \vert^2 \big) \qquad \partial_s f_s = \frac{\partial f}{\partial \vert \Psi \vert^2} \bigg\vert_{\vert \Psi_s \vert^2} 
\end{equation}

%\bibliography{}
\bibliography{Refs_experiment,Refs_theory}

\end{document}